\documentclass[conference,10pt]{IEEEtran}
\IEEEoverridecommandlockouts
\usepackage{tikz}
\usetikzlibrary{calc}
\usepackage{caption}
\usepackage{subcaption}
\usepackage{amsthm}
\usepackage{cite}
\usepackage{amsmath,amssymb,amsfonts}
\usepackage{mathtools}
\usepackage{algorithmic}
\usepackage{graphicx}
\usepackage{textcomp}
\usepackage{xcolor}
\usepackage{bbold}
\usepackage{mathrsfs}
\usepackage{qsymbols}
\usepackage[ansinew]{inputenc}
\usepackage{geometry}
   \usepackage{titlesec}
\usetikzlibrary{decorations.pathreplacing,calligraphy,backgrounds}
\def\BibTeX{{\rm B\kern-.05em{\sc i\kern-.025em b}\kern-.08em T\kern-.1667em\lower.7ex\hbox{E}\kern-.125emX}}

\begin{document}

\title{Coding for the Gaussian Channel in the Finite Blocklength Regime Using a CNN-Autoencoder}
\author{
\IEEEauthorblockN{Nourhan Hesham$^1$, Mohamed Bouzid$^2$, Ahmad Abdel-Qader$^1$, Anas Chaaban$^1$}
\IEEEauthorblockA{$^{1}$ School of Engineering, University of British Columbia, Kelowna, BC V1V1V7, Canada\\
$^{2}$ \'Ecole Polytechnique de Tunisie, Tunisia\\
Email:  \{nourhan.soliman,ahmad.abdelqader,anas.chaaban\}@ubc.ca, mohamed.bouzid@ept.u-carthage.tn}

\thanks{%
N. Hesham is on leave from the Department of Electronics and Electrical Engineering, Cairo University, Cairo, Egypt.
}
}


\maketitle

\begin{abstract}
The development of delay-sensitive applications that require ultra high reliability created an additional challenge for wireless networks. This led to Ultra-Reliable Low-Latency Communications, as a use case that 5G and beyond 5G systems must support. However, supporting low latency communications requires the use of short codes, while attaining vanishing frame error probability (FEP) requires long codes. Thus, developing codes for the finite blocklength regime (FBR) achieving certain reliability requirements is necessary. This paper investigates the potential of Convolutional Neural Networks autoencoders (CNN-AE) in approaching the theoretical maximum achievable rate over a Gaussian channel for a range of signal-to-noise ratios at a fixed blocklength and target FEP, which is a different perspective compared to existing works that explore the use of CNNs from bit-error and symbol-error rate perspectives. We explain the studied CNN-AE architecture, evaluate it numerically, and compare it to the theoretical maximum achievable rate and the achievable rates of polar coded quadrature amplitude modulation (QAM), Reed-Muller coded QAM, multilevel polar coded modulation, and a TurboAE-MOD scheme from the literature. Numerical results show that the CNN-AE outperforms these benchmark schemes and approaches the theoretical maximum rate, demonstrating the capability of CNN-AEs in learning good codes for delay-constrained applications.
\end{abstract}


\section{Introduction}
Over the last two decades, wireless communications have become a crucial part of our daily lives and have experienced exponential growth. Before the apparition of fifth-generation (5G) networks, these systems served human-centric networks such as conventional multimedia or voice-based services (images, videos, audio, etc.). However, the emergence of various services and applications (machine-type communication, remote medical diagnosis, smart cities, etc.) has imposed new constraints on wireless networks. For instance, some emerging applications have tight delay constraints and require high reliability to operate efficiently and safely (e.g. autonomous vehicles). For such applications, short codes must be used since they
contribute to low latency \cite{10}, and codewords should be chosen optimally to attain high reliability. 

From an information theory perspective, the existence of a codebook that can achieve a target frame error probability (FEP) at a given codelength is proven as long as the coding rate is below the theoretical maximum coding rate in the finite blocklength regime (FBR) characterized in \cite{1}. The challenge is to find this optimal code. From a practical perspective, conventional communication systems are often designed as a concatenation of independent blocks using a binary encoder/decoder and a modulator/demodulator to construct a code that is easy to analyze, optimize, and design. For example, the 5G in 3GPP Release 16 Specification \# 38.212 \cite{3gpp1} identifies polar codes \cite{polarArikan,polacodes2} for relatively long blocklength and Reed-Muller for short blocklength, combined QAM. Such design does not guarantee optimality compared to the theoretical maximum in the FBR.

To enhance the performance of conventional communication systems, joint design and optimization of coding and modulation in the FBR is necessary. 
Although this is often a tedious task, multilevel polar coded modulation (MLPCM) was introduced as a joint coding and modulation scheme that performs better than polar codes for Gaussian channels~\cite{ SCD,5G_MLPCM_1}, but unfortunately does not achieve the maximum achievable rates under short code-length. These limitations motivated the investigation of neural networks (NN) for code design in communication systems, after their success in various fields.


NNs have been shown to be a convenient tool for code design in communication systems, yielding better performance than conventional schemes in terms of error rates. Work in this area started by replacing one or more blocks of the physical layer with NNs for performance improvements, such as neural decoders \cite{2} and neural modulators and demodulators \cite{5,6,AE4}. Then, it was extended to propose an end-to-end (E2E) autoencoder (AE) based communication system in {\cite{3,4,AE3}}. More advanced NNs were proposed in~\cite{AE1} where a scheme using a Turbo-AE together with a feed-forward neural network (FFNN) for modulation is proposed. Yet, these methods only considered either specific coding rates (like a rate $1/3$ turbo code in \cite{AE1}), encoding a small number of symbols (ranging from $1$ to $15$), or a moderate number of information bits due to the complexity of the NN. Moreover, only the bit error rate (BER) of these NN was investigated. As such, the following question arises: Can an NN be used to design a code that approaches/achieves the information-theoretic maximum achievable rate?

To answer this question, one has to study the achievable information rate under a constraint on the FEP. In\cite{AE2}, the authors investigated the FEP of an end-to-end communication system where the coding and modulation are designed using an FFNN, with a limited number of symbols (up to $15$). However, \cite{AE2} did not investigate the maximum achievable rate of this architecture. This also applies to \cite{2,5,6,AE4,3,4,AE3,AE1}, it is currently unknown if any of the NNs proposed in these works achieves good performance compared to the maximum rate in the FBR, since this perspective was not studied in these papers. Complexity is also often a challenge in this area, for instance, training an FFNN-AE like in \cite{AE2} poses a challenge in terms of complexity and makes it difficult to study the performance under a larger number of symbols (say hundreds).


Since CNNs are known to be easier to train than many other NNs, in this paper, we investigate the potential of using a CNN-AE for coding and modulation in delay-sensitive applications, toward approaching the theoretical maximum rate in the FBR. To this end, we propose a CNN-AE architecture with few hyper-parameters and explain how to choose these parameters. Then, we jointly optimize the transmitter and receiver by training the AE, thus overcoming the mathematical complexity of finding an optimal code. Then, we compare the results with the performance of a scheme that combines polar codes and quadrature amplitude modulation (QAM), a scheme that combines Reed-Muller codes with QAM, a scheme that uses MLPCM \cite{5G_MLPCM_1}, and the turbo AE-based scheme proposed in \cite{AE1}. Results show that the proposed CNN-AE approaches the maximum coding rate in \cite{1} and outperforms the benchmark schemes, which highlights the capability of CNN-AEs in finding a good code that approaches the theoretical maximum rate in the FBR. Finally, we comment on the complexity and performance of similar designs using recurrent neural networks (RNN) or an FFNN instead of the CNN.

Next, we present the system model and review the theoretical maximum achievable rates in the FBR. Then, in Sec.~\ref{sec:Architecture}, we present our proposed CNN-AE architecture. Numerical results are presented and compared with results in~{\cite{1,polacodes2,5G_MLPCM_1,AE1}} in Sec.~\ref{sec:Results}. Finally, the paper is concluded and future work directions are presented in Sec.~\ref{sec:Conclusion}.

\section{System Model}\label{sec:System_Model}
For a general communication system, the transmitter feeds a message $\textbf{s}$ with $K$ information bits to an encoder that encodes the bits, with rate $R$, to $\mathbf{x}=(x_1,x_2,\ldots,x_n)\in\mathbb{C}^n$ which must satisfy a power constraint 
\begin{equation}\label{eq:constraint}
    \frac{1}{n}\sum_{i=1}^n|x_i|^2\leq P.
\end{equation} 
 Then, $\mathbf{x}$ is sent through a Gaussian channel leading to the received signal
$$\textbf{y}=\textbf{x}+\textbf{w}$$
where $\textbf{w}$ is an $n$-dimensional vector of independent and identically distributed circularly symmetric complex Gaussian noise with zero mean and variance $N_0$. 

At the receiver side, the decoder is designed to find an estimate $\hat{\textbf{s}}$ of the transmitted message, using the conditional probability of the channel $\mathbb{P}_{\textbf{Y} | \textbf{X}}(\textbf{y}| \textbf{x})$, such that the FEP $\mathbb{P}(\hat{\textbf{s}} \ne {\textbf{s}})$ does not exceed a predetermined value $\varepsilon$. The trade-off between the blocklength $n$ and
the FEP  $\varepsilon$ forms a limitation on the performance of the system in terms of the achievable information rate. Note that latency and reliability depend on $n$ and $\varepsilon$, respectively. 

\begin{centering}
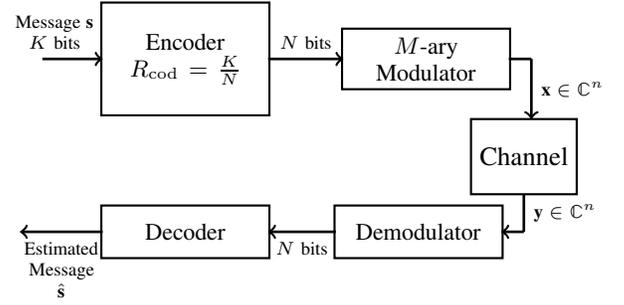
\begin{figure}
\centering
\begin{tikzpicture}[thick,scale=1, every node/.style={scale=1}]
\draw [->,line width=1] (-5 ,2) -- node[pos=0.2,text width=2cm,align=center,above,thick,font=\scriptsize] {Message $\textbf{s}$\\     $K$ bits  
}(-4.2,2) ;
\node (demux) at (-3.1,2) [draw,align=center,text width=2cm, thick,minimum width=0.1cm,minimum height=1.5cm, font=\small] {Encoder\\ $R_{\rm cod}=\frac{K}{N}$} ;

\draw [->,line width=1] (-2,2) -- node[above,font=\scriptsize] {$N $ bits}(-1,2) ;

\node (Mapper) at (0.1,2) [draw,thick,align=center,text width=2cm, minimum width=1cm,minimum height=0.7cm, font=\small] {$M$-ary Modulator } ;

\draw [-,line width=1] (1.2 ,2) --  (1.5,2);
\draw [->,line width=1] (1.5 ,2) --  node[pos=0.5,right,font=\scriptsize] {$\textbf{x}\in \mathbb{C}^n$} (1.5,1.2);

\node (channel) at (1.4,0.7) [draw,thick,align=center,minimum width=1cm,minimum height=1cm, font=\normalsize] {Channel} ;

\draw [-,line width=1] (1.4 ,0.2) --  node[pos=0.5,right,font=\scriptsize] {$\textbf{y}\in \mathbb{C}^n$} (1.4,-0.3);
\draw [->,line width=1] (1.4 ,-0.3) --  (1.1,-0.3);

\node (Mapper) at (-0,-0.3) [draw,thick,align=center,text width=2cm, minimum width=1cm,minimum height=0.7cm, font=\small] {Demodulator } ;

\draw [->,line width=1] (-1.1 ,-0.3) --  node[pos=0.5,below,font=\scriptsize] {$N$ bits} (-2,-0.3);

\node (demux) at (-3.1,-0.3) [draw,align=center,text width=2cm, thick,minimum width=0.1cm,minimum height=0.7cm, font=\small] {Decoder} ;

\draw [->,line width=1] (-4.2 ,-0.3) --  node[pos=0.5,text width=2cm, align=center, below,font=\scriptsize] {Estimated\\ Message\\ $\hat{\textbf{s}}$ } (-5.3,-0.3);
\end{tikzpicture}
\caption{A conventional communication system consisting of an encoder, modulator, demodulator, and decoder.}
\label{fig:CS}
\end{figure}
\end{centering}

For a conventional communication system (Fig. \ref{fig:CS}), the encoding process is divided into channel coding with coding rate $R_{\rm cod}= \frac{K}{N}$ where the encoder maps the $K$ information bits to a codeword of length $N$ bits, followed by an $M$-ary modulator of order $M=2^{k_{\rm mod}}$ which maps each $k_{\rm mod}$ coded bits into a complex-valued symbol to obtain the vector $\mathbf{x}$ of length $n$, where $n=N/k_{\rm mod}$. The overall rate of information transmission is given by $R= R_{\rm cod} k_{\rm mod}$ bits per complex-valued transmission. This model will serve as one of the benchmarks we use for comparison.

The achievable rate of the conventional system described above usually deviates from the channel capacity (the maximum achievable rate) due to the separation of the design of channel coding and modulation. This capacity was first characterized by Shannon in~\cite{7}, as the maximum rate of information transmission such that $\varepsilon\to0$ as $n \rightarrow \infty$. For a Gaussian channel with signal-to-noise  (SNR) $$\gamma=\frac{P}{N_0},$$ the capacity is given by
\begin{equation}\label{eq:Capacity}
C=\log_{2}(1+\gamma)\quad \text { bits/transmission},
\end{equation}
which can be achieved using a code that maps the information bits s into the complex-valued codeword $\mathbf{x}$ without separating this mapping into binary coding and modulation. However, for a delay-constrained application, short codes (small $n$) are used to ensure low latency. In this case, Shannon's capacity which is defined for infinite blocklength $n$ and vanishing error probability $\varepsilon$  becomes inaccurate. Polyanskiy {\it et al.} \cite{1} derived a closed-form approximation of the maximum achievable rate for the {Gaussian} channel in the FBR, for a blocklength $n$ as short as $100$ symbols and target FEP  $ 0 <  \varepsilon < 0.5 $. This maximum rate is characterized for a Gaussian channel with SNR $\gamma$ as follows
\begin{equation}\label{eq:Rate}
R_{n,\varepsilon} = C- \sqrt{\frac{V}{n}} Q^{-1}(\varepsilon) + O(\log n),
\end{equation}
where $C$ is defined in \eqref{eq:Capacity}, $Q^{-1}(\cdot)$ is inverse Q-function, $O(\log(n))$ is bounded in \cite{1} by \textcolor{black}{$O(1)<O\left(\log_2(n)\right)<\frac{1}{2n}\log_2(n)+O( 1)$} and we neglect the term $O( 1)$ in our analysis and use the upper bound, and  $$V= \frac{\gamma(\gamma+2)}{(\gamma+1)^{2}} \log_2 ^{2} e$$ is the channel dispersion defined as the variance of the mutual information density between $x_i$ and $y_i$.

The maximum achievable rate is defined as the capacity with a back-off term which is a function of the blocklength $n$, FEP $\varepsilon$, and SNR. This theoretical result will be used as another benchmark for assessing the performance of our proposed scheme.

The separation of channel coding and modulation in the conventional communication system shown in Fig. \ref{fig:CS} leads to a sub-optimal system. This simplification is often done because finding a coding scheme for the general system which approaches \eqref{eq:Rate} is generally cumbersome. The objective of this paper is to propose an NN to jointly optimize the transmitter and receiver functions to outperform conventional methods and approaches \eqref{eq:Rate}, as described next. 
 
\begin{figure}
\centering
\begin{tikzpicture}[thick,scale=1, every node/.style={scale=1}]
\node (Input) at (0,0) [draw,thick,align=center,text width=1cm, minimum width=2cm,minimum height=0.5cm,rounded corners,fill=blue!15, font=\small] {Input } ;

\draw [->,line width=1.5] (0 ,-0.3) --  node[pos=0.5,right,font=\small] {\footnotesize $K$ bits} (0,-0.7);

\draw [decorate,decoration={brace,amplitude=10pt},xshift=-4pt,yshift=-2pt]
(1.3,-0.5) -- (1.3,-3.5) node [black,midway,xshift=1cm] 
{\footnotesize Encoding};


\node (Conv1D_1) at (0,-1) [draw,thick,align=center,text width=1cm, minimum width=2cm,minimum height=0.5cm,rounded corners, font=\small] { Conv1D} ;

\node (Conv1D_2) at (0,-2) [draw,thick,align=center,text width=1cm, minimum width=2cm,minimum height=0.5cm,rounded corners,fill=white, font=\small] { Conv1D} ;

\node (Conv1D_3) at (0,-3) [draw,thick,align=center,text width=1cm, minimum width=2cm,minimum height=0.5cm,rounded corners,fill=white, font=\small] { Conv1D} ;
\draw [->,line width=1.5] (0 ,-1.25) --  node[pos=0.5,left,font=\scriptsize] {$L\times M_1$} (0,-1.75);

\draw [->,line width=1.5] (0 ,-2.25) --  node[pos=0.5,left,font=\scriptsize] {$L\times M_1$} (0,-2.75);

\node (Reshape Layer) at (0,-4.1) [draw,thick,align=center,text width=1cm, minimum width=2cm,minimum height=0.5cm,rounded corners,fill=blue!15, font=\small] {Reshape Layer } ;

\draw [->,line width=1.5] (0 ,-3.25) --  node [pos=0.5,left,font=\scriptsize] {$L\times N'$} (0,-3.7);

\draw [decorate,decoration={brace,amplitude=10pt},xshift=-4pt,yshift=-2pt] (1.4,-4.6) -- (1.4,-6.6) node [black,midway,xshift=1.1cm]  {\footnotesize Modulation};


\node (Conv1D_4) at (0,-5.25) [draw,thick,align=center,text width=1cm, minimum width=2cm,minimum height=0.5cm,rounded corners,fill=white, font=\small] {Conv1D } ;

\node (Conv1D_5) at (0,-6.25) [draw,thick,align=center,text width=1cm, minimum width=2cm,minimum height=0.5cm,rounded corners,fill=white, font=\small] { Conv1D} ;

\draw [decorate,decoration={brace,amplitude=10pt},xshift=-4pt,yshift=-2pt]
(5,-7.5) -- (5,-5.5) node [black,midway,xshift=-1.25cm] 
{\footnotesize Demodulation};
\draw [->,line width=1.5] (0 ,-4.5) --  node[pos=0.5,left,font=\scriptsize] {$n\times k_{mod}$} (0,-5);

\draw [->,line width=1.5] (0 ,-5.5) --  node[pos=0.5,left,font=\scriptsize] {$n\times M_2$} (0,-6);


\draw [->,line width=1.5] (0 ,-6.5) --  node[pos=0.55,right,align=left,font=\scriptsize,text width=1.8cm] {$n$  complex-valued symbols} (0,-7);
\node (Normalization Layer) at (0,-7.5) [draw,thick,align=center,text width=2cm, minimum width=2cm,minimum height=0.5cm,rounded corners,fill=blue!15, font=\small] {Normalization Layer } ;

\draw [-,line width=1.5] (0 ,-7.95) --  node[pos=0.5,left,font=\small] {} (0,-8.25);

\draw [->,line width=1.5] (0,-8.25) --  node[pos=0.5,below,align=center,font=\small,text width=1.5cm] {$n$ complex-valued symbols} (1.86,-8.25);

\node (Channel) at (3,-8.25) [draw,thick,align=center,text width=2cm, minimum width=1.5cm,minimum height=0.5cm,rounded corners,fill=blue!15, font=\small] {Channel } ;

\draw [-,line width=1.5] (4.1 ,-8.25) --  node[pos=0.5,below,align=center,font=\small,text width=1.5cm] {$n$ complex-valued symbols} (5.8,-8.25);

\draw [->,line width=1.5] (5.8 ,-8.25) --  node[pos=0.28,below,font=\scriptsize] {} (5.8,-7.5);


\node (Conv1D_6) at (5.8,-6) [draw,thick,align=center,text width=1cm, minimum width=2cm,minimum height=0.5cm,rounded corners,fill=white, font=\small] { Conv1D} ;
\draw [->,line width=1.5] (5.8 ,-7.5) --  node[pos=0.7,right,font=\scriptsize] {$n\times M_2$} (5.8,-6.25);

\node (Conv1D_7) at (5.8,-7.2) [draw,thick,align=center,text width=1cm, minimum width=2cm,minimum height=0.5cm,rounded corners,fill=white, font=\small] { Conv1D} ;
\draw [->,line width=1.5] (5.8 ,-5.75) --  node[pos=0.5,right,font=\scriptsize] {$n\times k_{mod}$} (5.8,-5.25);

\node (Reshape Layer) at (5.8,-4.8) [draw,thick,align=center,text width=1cm, minimum width=2cm,minimum height=0.5cm,rounded corners,fill=blue!15, font=\small] {Reshape Layer } ;

\draw [->,line width=1.5] (5.8 ,-4.35) --  node[pos=0.5,right,font=\scriptsize] {$L\times N'$} (5.8,-3.3);

\draw [decorate,decoration={brace,amplitude=10pt},xshift=-4pt,yshift=-2pt]
(5,-3.3) -- (5,-0.5) node [black,midway,xshift=-1cm] 
{\footnotesize Decoding};

\draw [->,line width=1.5] (5.8 ,-3.35) --  node[pos=0.8,right,font=\scriptsize] {$L\times M_1$} (5.8,-2.25);
\draw [->,line width=1.5] (5.8 ,-2.35) --  node[pos=0.8,right,font=\scriptsize] {$L\times M_1$} (5.8,-1.3);

\node (Conv1D_8) at (5.8,-1) [draw,thick,align=center,text width=1cm, minimum width=2cm,minimum height=0.5cm,rounded corners, font=\small] {Conv1D } ;

\node (Conv1D_9) at (5.8,-2) [draw,thick,align=center,text width=1cm, minimum width=2cm,minimum height=0.5cm,rounded corners,fill=white, font=\small] {Conv1D } ;

\node (Conv1D_10) at (5.8,-3) [draw,thick,align=center,text width=1cm, minimum width=2cm,minimum height=0.5cm,rounded corners,fill=white, font=\small] { Conv1D} ;

\draw [->,line width=1.5] (5.8 ,-0.75) --  node[pos=0.4,right,font=\small] {\footnotesize $K$ bits} (5.8,-0.25);

\node (Output) at (5.8,0) [draw,thick,align=center,text width=1cm, minimum width=2cm,minimum height=0.5cm,rounded corners,fill=blue!15, font=\small] {Output } ;

\end{tikzpicture}
\caption{Proposed CNN-AE. The `encoding', `modulation',
`demodulation', and `decoding' layers are trained jointly.}
\label{fig:Architecture}
\end{figure}
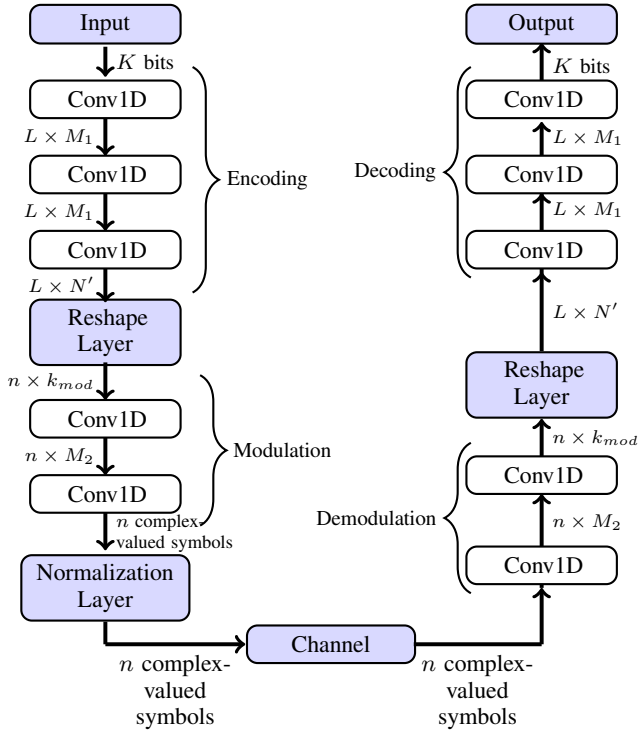

\section{Autoencoder for the Finite Blocklength Regime}\label{sec:Architecture}
\subsection{Neural Network structure}
In this section, a CNN-AE that is capable of approaching the maximum rate in the FBR is proposed. The structure of our CNN-AE is shown in Fig.~\ref{fig:Architecture}. Different from \cite{5}, encoding and decoding blocks are added around the modulator and the demodulator. In our approach, we model the $4$ blocks  (channel encoder, modulator, demodulator, and channel decoder) as NNs and train them jointly in an end-to-end manner, while using $K$ bits as an input at the transmitter (output at the receiver) and $n$ complex-valued symbols as an output at the transmitter (input at the receiver). The AE's objective is to find an $n$-dimensional code with rate $R$ bits/transmission that allows the reconstruction of the message at the receiver side with an FEP less than $\varepsilon$. Thus, the AE aims to learn the codebook in an end-to-end manner based on the channel rather than the conventional method of using a careful mathematical construction of encoding/decoding and modulation/demodulation. 

The CNN-AE architecture is designed to mimic the blocks of a conventional communication system. Hence, each block is represented by a set of layers, and its dimensions are chosen depending on the function it performs. Let us suppose that we want to transmit a message with $K$ information bits through an encoder mapping these bits into $N$ coded bits at rate $R_{\rm cod}=K/N$, followed by a modulator that maps the $N$ bits into $n$ complex-valued symbols using a modulation order $2^{k_{\rm mod}}$ where $k_{\rm mod}=N/n$. Since the code rate is $R_{\rm cod}=K/N$, to simplify this encoding into a code with a lower number of input/output bits, we write $K=K'L$ and $N=N'L$ where $L$ is the greatest common divisor of $N$ and $K$. This allows us to interpret the encoding of $K$ bits into $N$ bits at rate $R_{\rm cod}$ as the encoding of $L$ sub-messages of $K'$ bits each into $L$ sub-codewords of $N'$ bits each at rate $R_{\rm cod}$. Thus, the input of the AE will be $K=K'L$ bits. Then to construct the codewords, we use three $1$-D convolutional (Conv1D) \textcolor{black}{layers}.\footnote{\textcolor{black}{A larger number of layers can be used if needed, but it was noticed that 3 layers suffice for the purpose of this paper.}}. The first two Conv1D layers will map each sub-message of $K'$ information bits ($L$ sub-messages) into a higher dimensional space of $M_1$ dimensions, \textcolor{black}{leading to} an output size of $L\times M_1$. This will allow the AE to learn a good placement of each of the $L$ sub-codewords in an $M_1$ dimensional space. Then, the third Conv1D layer is used to map the sub-codewords down to an $N'$ dimensional space (output size $L\times N'$). Finally, to prepare the ${LN'}=N$ symbols for modulation into $n$ complex-valued symbols, they are reshaped into an $n\times k_{\rm mod}$ matrix.

To modulate the $n$ symbols with $k_{\rm mod}$ symbols each, we use two Conv1D layers. The first Conv1D layer maps the $k_{\rm mod}$ symbols into a higher dimensional space of $M_2$ dimensions using a Conv1D layer (output size $n\times M_2$). This allows the AE to learn a good placement of the modulation symbols in an $M_2$-dimensional space. Then, the other Conv1D layer maps each of the $n$ modulated $M_2$-dimensional symbols into a complex-valued symbol which we represent using a $2$-dimensional real-valued representation via another Conv1D layer (output size $n\times 2$). A (nontrainable) normalization layer is finally added to satisfy the average power constraint in \eqref{eq:constraint} by normalizing by the average power of the $n$ symbols. Then, after the (nontrainable) channel layer, the receiver side including the demodulator and decoder are designed in the same manner to reconstruct the transmitted message. The parameters of AE are summarized in Table \ref{tab:Parameters}.


We use Conv1D layers due to their lower complexity and better trainability than FFNN and RNN \cite{11}. Also, each layer is followed by an Exponential Linear Unit (ELU) activation function to allow for non-linear coding. However, we use a linear activation function at the end of the mapper to allow for any positive or negative values at the output of the modulation. In the last layer at the receiver side, we use a sigmoid activation function followed by thresholding at $0.5$ to ensure that the output is a binary vector. Finally, we note that each trainable layer is followed by a Batch Normalization layer to help the model converge quickly.


\subsection{Methodology}\label{sec:Methodology}
Since this work aims to demonstrate the existence of an AE that can achieve a rate close to the maximum achievable rate in the FBL, we fix a blocklength $n$ and a target FEP $\varepsilon$. Then, we train the AE in an end-to-end manner at a certain SNR $\gamma$ and rate $R$, by feeding a dataset of information bits ($K$ bits in each data sample) and training via backpropagation. Then, we test the AE using a different set of data and evaluate its FEP. If the FEP is larger than $\varepsilon$, we gradually decrease $R$ via choosing $R_{\rm cod}$ and $k_{\rm mod}$ until the FEP becomes smaller than $\varepsilon$. The CNN parameters are chosen accordingly, based on the parameters presented in the previous section, where $R_{\rm cod}$ and $k_{\rm mod}$ are selected so that $R_{\rm cod}k_{\rm mod}=R$, while the parameters $M_1$ and $M_2$ (whose impact is analyzed in Sec IV) are chosen as $200$ and $100$, respectively, since they turned out to work well for the range of rates, code-length, and FEP considered in this study.

It is worth noting that works in the literature which study the BER achieved by an AE \cite{AE1,2,3,4,5} train the AE at a certain SNR with a fixed rate $R$ and test it at a wide range of SNRs where it can be observed that the BER decreases as SNR increases. This can also be done in our work. However, the goal of our work is to show that at any SNR, the proposed CNN-AE architecture can approach the maximum rate in the FBR. Since this maximum rate increases with SNR, so must the rate of the CNN-AE. Since increasing the rate of the CNN-AE requires changing the dimensions of its input and output, this implies that a new training is needed when the SNR is increased. Note that this is not unique to our CNN-AE, since increasing the rate of a conventional scheme requires changing the coding and/or modulation parameters which is analogous to changing the architecture in our case.
\begin{table}[t]
\centering
\begin{tabular}{|c|c|c|} 
\hline 
Type of layer         & Activation function             & Output size            \\ 
\hline
\multicolumn{1}{|l}{} & \multicolumn{1}{c}{\textbf{Encoder}}     & \multicolumn{1}{l|}{}  \\ 
\hline
Input                 & None                            & ($LK'\times1$)               \\ 
\hline
Conv 1D               & elu                             & (${L}\times M_1$)               \\ 
\hline
Conv 1D               & elu                             & (${L}\times M_1$)               \\ 
\hline
Conv 1D               & elu                             & (${L}\times N^\prime$)                 \\ 
\hline
Reshape               & None                            & ($n\times k_{mod}$)        \\ 
\hline
\multicolumn{1}{|c}{} & \multicolumn{1}{c}{\textbf{Modulator}}   & \multicolumn{1}{l|}{}  \\ 
\hline
Conv1D                & elu                             & ($n\times M_2$)               \\ 
\hline
Conv1D                & linear                          & ($n\times 2$)                \\ 
\hline
Normalization Layer   & None                            & ($n\times 2$)                \\ 
\hline
\multicolumn{1}{|l}{} & \multicolumn{1}{c}{\textbf{Channel}}     & \multicolumn{1}{l|}{}  \\ 
\hline
{Gaussian} channel          & None                            & ($n\times 2$)                \\ 
\hline
\multicolumn{1}{|l}{} & \multicolumn{1}{c}{\textbf{Demodulator}} & \multicolumn{1}{l|}{}  \\ 
\hline
Conv1D                & elu                             & ($n\times M_2$)               \\ 
\hline
Conv1D                & linear                          & ($n\times k_{mod}$)       \\ 
\hline
Reshape               & None                            & (${L}\times N^\prime$)                 \\ 
\hline
\multicolumn{1}{|l}{} & \multicolumn{1}{c}{\textbf{Decoder}}     & \multicolumn{1}{l|}{}  \\ 
\hline
Conv1D                & elu                             & (${L}\times M_1$)               \\ 
\hline
Conv1D                & elu                             & (${L}\times M_1$)               \\ 
\hline
Conv1D                & sigmoid                         & (${L}K'\times 1$)                \\
\hline
\end{tabular}
\caption[Caption for LOF]{Parameters of the proposed CNN-AE.}
\label{tab:Parameters}
\end{table}

\section{Numerical Results}\label{sec:Results}
In this section, simulations are performed to find the maximum rates that can be achieved at given FEP $\varepsilon$ and blocklength $n$. Numerical results are presented based on the methodology provided in Sec. \ref{sec:Methodology} under the settings provided next. The AE performance is compared to the performance of polar codes combined with QAM, Reed-Muller codes combined with QAM, MLPCM, and the TurboAE-MOD scheme in~\cite{AE1} in addition to the theoretical maximum provided in \eqref{eq:Rate}.

\subsection{Simulation parameters}
For simulations, we train and test the CNN-AE under a Gaussian channel at blocklength $n=128$ and FEP $\varepsilon=10^{-2}$. We generate $10^6$ random binary vectors for each of the training and the testing sets. The training is conducted using Adam optimizer \cite{Adam} with a learning rate of $0.001$ and binary cross entropy loss, while the parameters $M_1$ and $M_2$ are chosen as $200$ and $100$, respectively. The batch size is set to $500$ while the number of epochs is $100$. Finally, the training SNR, a communication-related parameter, was fixed at a value that matches (equal or slightly lower than) the testing SNR. Simulation results are discussed next.
   \begin{table}[t]
\centering
\begin{tabular}{|c|c|c|c|} 
\hline\hline
$M_1$,$M_2$ & $(50,20)$ & $(80,50)$ & $(200,100)$ \\ 
\hline
Rate $3$ bits/transm.  & $\times$  & $\times$ & $\checkmark$  \\ 
\hline
Rate $1$ bits/transm. & $\checkmark$  & $\checkmark$ & $\checkmark$  \\ 
\hline
\end{tabular}
\caption{Performance of the AE at different values of $M_1$ and $M_2$ and $\text{SNR}=10\ \text{dB}$. Here, `$\times$' means that the target FEP was not achieved, whereas `$\checkmark$' means it was achieved.}
\label{tab:filters}
\end{table}

\subsection{Results}

\begin{figure}[t]
            \centering
\includegraphics[width=1\linewidth,height=7.5cm]{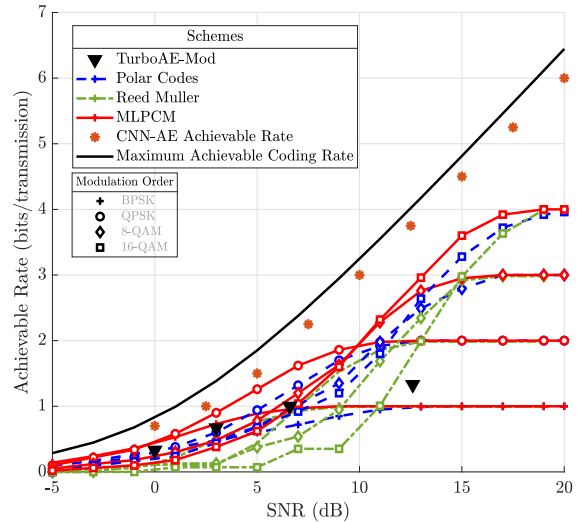}            \caption{The simulated achievable rates of the CNN-AE compared with schemes from the literature and theoretical achievable rates}
            \label{fig:AIR}
        \end{figure}

The achievable rates that satisfy an FEP $< 10^{-2}$ for a blocklength of $n=128$ are presented in Fig.~\ref{fig:AIR}, in addition to the maximum rate in the FBR, and the achievable rate polar and Reed-Muller coded QAM (with hard detection), MLPCM, and the AE-based scheme in \cite{AE1}. While these schemes are conventional and employ separate encoders and modulators, the MLPCM scheme uses joint coding and modulation where the encoding and modulation are jointly designed so that each symbol carries information from different polar codes (layers) and the demodulation and decoding are done in a hierarchical manner (multi-stage decoding) \cite{5G_MLPCM_1}. {The benchmark AE-based scheme is the TurboAE-MOD scheme provided in \cite{AE1} which jointly trains a Turbo-AE and a QAM modulator/demodulator, and whose performance is comparable to a classical turbo code combined with a QAM modulator as shown in \cite{AE1}}. 

It can be seen from the figure that the CNN-AE outperforms the benchmark schemes, and achieves rates close to the theoretical maximum provided in \eqref{eq:Rate}. This demonstrates the capability of the proposed CNN-AE in finding a good
code for the channel in the FBR. It is worth noting that the TurboAE-MOD scheme, in \cite{AE1} has the limitation of a fixed encoding rate of $1/3$, which makes its architecture less flexible. Thus, this TurboAE-MOD scheme provides rates that are multiples of $1/3$ depending on the selected modulation order.\footnote{Note that the rate of the scheme in \cite{10} can be improved using an AE architecture which achieves a higher coding rate, which is worth additional investigation. However, our proposed CNN-AE demonstrates that there is no need to restrict attention to a TurboAE-MOD architecture.} This rate control is coarse compared to the proposed CNN-AE which has more flexibility in choosing the coding rate and modulation order so that it approaches the theoretical bound.

The dimensions of the encoder and modulator filter space ($M_1$ and $M_2$) were further investigated in Table.~\ref{tab:filters}. The table records the ability of the network to reach the rates of $3$ and $1$ bits per transmission at $\text{SNR}=10\ \text{dB}$ for different $M_1$ and $M_2$. At higher rates closer to the maximum achievable rate ($R=3$), it was found that small values of $M_1$ and $M_2$ constrain the performance of the AE which makes it unable to reach the target FEP. However, at lower rates ($R=1$), it was found that smaller $M_1$ and $M_2$ can achieve the desired performance. This is because larger $M_1$ and $M_2$ give the AE more flexibility in placing the codewords and modulation symbols in a higher dimensional space before mapping back to $n$ symbols. However, it must be noted that larger $M_1$ and $M_2$ increase the training complexity.


Finally, during our experiments, we investigated different types of AEs (FFNN, RNN, CNN) to examine their complexity. We found that to achieve a rate of $R=2$ for instance at $n=128$, $R_{\rm cod}=\frac{1}{2}$, $k_{\rm mod}=4$, and $\varepsilon=10^{-2}$ for $\text{SNR}=12\ dB$, the RNN needs $252,427$ trainable parameters while the CNN needs $103,861$ trainable parameters at $M_1=100$ and $M_2=20$. Moreover, the FFNN needed $723,200$ trainable parameters and was still unable to achieve this rate making it inefficient and nonscalable to larger blocklengths. Comparing the RNN and the CNN, we found that they achieved similar performance in terms of achievable rates, but the CNN-AE was faster in terms of training time thanks to its parameter sharing technique. Hence, the proposed CNN-AE provides a better performance in terms of coding rate and complexity.

\section{Conclusion}\label{sec:Conclusion}
We proposed a CNN-AE that provides achievable rates over the Gaussian channel which approaches the theoretical maximum rate  in the FBR. We also showed that the proposed CNN-AE outperforms benchmark conventional schemes and another NN-based AE, the TurboAE-MOD scheme in \cite{AE1}. The main conclusion of this work is that a CNN-AE can be used to find a code for the Gaussian channel in the FBR which approaches the theoretical maximum rate. This work can be extended to cover more complex channels that exist in real-world scenarios such as interference channels and Rayleigh fading channels. Furthermore, with the increasing demand and the apparition of 5G networks with dense deployment and massive connectivity, a promising research direction is to consider the performance of CNN-AE in large-scale wireless networks in the FBR.

\bibliographystyle{IEEEtran}
\bibliography{references}

\end{document}